\documentclass{article}
\usepackage{graphicx} 

\usepackage[utf8]{inputenc}
\usepackage{amsmath}
\usepackage{amsfonts}
\usepackage{amssymb}
\usepackage{graphicx}
\usepackage{url}
\usepackage{hyperref} 
\usepackage{listings} 
\usepackage{xcolor}   
\usepackage{geometry}
\usepackage{framed}
\usepackage{quoting}

\lstdefinestyle{mystyle}{
    language=SQL, 
    basicstyle=\ttfamily\footnotesize,
    keywordstyle=\color{blue},
    commentstyle=\color{green!40!black},
    stringstyle=\color{purple},
    breakatwhitespace=false,         
    breaklines=true,                 
    captionpos=b,                    
    keepspaces=true,                 
    numbers=left,                    
    numbersep=5pt,                  
    showspaces=false,                
    showstringspaces=false,
    showtabs=false,                  
    tabsize=2
}
\title{\textbf{miRKatAI: An Integrated Database and Multi-agent AI system for microRNA Research}}
\author{
Karen Guerrero-Vazquez\textsuperscript{1,2,3}, Jacopo Umberto Verga\textsuperscript{1,3},\\
Pilib \'{O} Broin\textsuperscript{1,2,3}, Eva Szegezdi\textsuperscript{3,4},\\
 Katarzyna Goljanek-Whysall\textsuperscript{2,3,5,6}\\
\normalsize \textsuperscript{1} School of Mathematical and Statistical Sciences, University of Galway, Galway, Ireland.\\
 \normalsize \textsuperscript{2} Galway RNA Research Cluster, University of Galway, Galway, Ireland.\\
 \normalsize \textsuperscript{3} Centre for Research Training in Genomics Data Science, University of Galway, Galway, Ireland.\\
\normalsize \textsuperscript{4} Apoptosis Research Centre, University of Galway, Galway, Ireland.\\
\normalsize \textsuperscript{5} School of Pharmacy and Medical Sciences, University of Galway, Galway, Ireland.\\
\normalsize \textsuperscript{6} Institute of Life Course and Medical Sciences, University of Liverpool, Liverpool, UK.\\
      Co-corresponding authors: \\
       \normalsize K.GuerreroVazquez1@universityofgalway.ie \\
        \normalsize Kasia.whysall@universityofgalway.ie
}

\date{} 

\colorlet{shadecolor}{lightgray}
\newenvironment{shadedquotation}
 {\begin{shaded*}
  \quoting[leftmargin=0pt, vskip=0pt]
 }
 {\endquoting
 \end{shaded*}
}

\begin{document}
\maketitle


\begin{abstract}
MicroRNAs (miRs) are robust regulators of gene expression, implicated in most biological processes. microRNAs predominantly downregulate the expression of genes post-transcriptionally and each miR is predicted to target several hundred genes. The accurate identification and annotation of miR–mRNA target interactions is central to understanding miRs function and their therapeutic potential. However, computational target prediction is challenging due to imperfect complementarity of miRs with their targets and the growing volume and heterogeneity of experimental data present challenges in accessing, integrating, and analysing miR–target interaction information across biological contexts. This creates a need for integrated resources and intelligent query tools. 

We present the miRKat Suite, comprising miRKatDB, a comprehensive, curated database of predicted and validated miR-target interactions and associated annotations, and miRKatAI, a multi-agent system powered by large language models (LLMs) and LangGraph. miRKatDB integrates data from multiple publicly available sources, providing a comprehensive foundation for miR studies, including miR target genes and changes in levels of tissue expression previously reported. miRKatAI offers a natural language interface for complex querying of miRKatDB, facilitates grounded information retrieval from established sources in the field, and supports basic data visualisation. The miRKat Suite aims to accelerate miR research by streamlining data access, enhancing exploratory analysis, and supporting hypothesis generation.

\end{abstract}


\section*{Background}\label{sec:introduction}

MicroRNA (miR) research plays a pivotal role in understanding complex biological processes and developing novel therapeutic strategies. These small non-coding RNAs are crucial regulators of gene expression \cite{Wu2018TheCancer}, implicated in multiple biological functions, from cellular development and physiology to complex pathologies such as cancer \cite{Wu2018TheCancer} and ageing \cite{Yanai2020MicrornasReview}. The rapid accumulation of diverse miR data, encompassing sequences, targets, expression profiles, and functional annotations, presents significant challenges for researchers in terms of data navigation and interpretation, in addition to reliable target prediction.

The miR information landscape is often fragmented, with critical data dispersed across multiple specialised databases. Accessing, integrating, and querying this structured information can demand advanced bioinformatics skills, creating a barrier for many bench scientists and clinicians \cite{Attwood2019ANeeds}. Beyond structured data, staying current with the expanding literature to understand miR functions, mechanisms, and context is a laborious and time-consuming \cite{Rodriguez-Esteban2009BiomedicalApplications}. Furthermore, visualising data trends, such as expression levels across tissues or the number of targets for specific miRs, frequently involves manual data processing using disparate software tools. Collectively, these challenges consume valuable time and resources, diverting focus from core scientific inquiry and hypothesis generation.

To address these challenges and streamline miR research, we developed the miRKat Suite. This integrated platform comprises two core components: miRKatDB and miRKatAI. miRKatDB is a custom-developed relational database that consolidates and harmonises miR-related information by integrating data from key public repositories, including miRBase, the state-of-the-art miR database \cite{Kozomara2019MiRBase:Function}, TargetScan with predicted miRs seeds based on conservation \cite{Agarwal2015PredictingMRNAs}, miRDB, a commonly used miR target prediction database \cite{Wang2008MiRDB:Interface}, and miRTarBase, which contains experimentally validated miR targets  \cite{Huang2022MiRTarBaseInteractions}. The original datasets were meticulously curated to align with the database format while preserving their authentic content. This structured approach establishes a robust foundation for miR studies, facilitating comprehensive analysis and interpretation.

miRKatAI is a multi-agent system, a concept demonstrating significant potential in biological applications \cite{Mehandru2025BioAgents:Systems}. It is constructed using Google's Gemini large language models \cite{Team2023Gemini:Models} and the LangGraph framework for a modular approach \cite{Wang2024AgentModels}.
miRKatAI serves as an interface to miRKatDB, enabling researchers to perform complex queries from multiple databases simultaneously using natural language. It facilitates the retrieval of practical information from external scientific literature and databases to augment database findings and supports the generation of basic data visualisations based on this retrieved information.

The miRKat Suite aims to accelerate miR research by simplifying data access, enhancing exploratory analysis, and supporting hypothesis generation. By lowering the barrier to complex data exploration and synthesis, the suite empowers researchers to navigate the miR landscape and uncover novel biological insights more efficiently.

\section*{Implementation}
\subsection*{miRKatDB}
\label{sec:mirkatdb}
miRKatDB is a custom-developed relational database designed to address the fragmentation of microRNA data. It consolidates and harmonises miR-related information from various public repositories into a single, queryable resource. By integrating diverse datasets, miRKatDB provides a structured and comprehensive foundation to support and streamline miR research. miRKatDB contains miRs' basic data, microRNA expression levels in different tissues, systems, and organs, microRNA-gene interaction and predicted seed of the mature microRNAs.


\subsubsection*{Data Sources and Content}\label{sec:mirkatdb_sources}

miRKatDB integrates data from several key public repositories to provide a comprehensive resource for miR research. The foundational data, including miR nomenclature and sequences, is primarily sourced from \textbf{miRBase} \cite{Kozomara2019MiRBase:Function, Griffiths-Jones2006MiRBase:Nomenclature}. Information regarding microRNA-gene interactions is curated from multiple databases, including \textbf{miRDB} \cite{Wang2008MiRDB:Interface} for predicted interactions and \textbf{miRTarBase} \cite{Huang2022MiRTarBaseInteractions} for experimentally validated interactions.

To ensure a comprehensive understanding of miR target recognition mechanisms, miRKatDB incorporates crucial seed region information from \textbf{TargetScan} \cite{Agarwal2015PredictingMRNAs}. Functional context for miRs is further enriched by integrating tissue-specific expression data from \textbf{miRNATissueAtlas2} \cite{Keller2022MiRNATissueAtlas2:Atlas}. Moreover, to guarantee accurate data mapping and seamless integration across diverse datasets, miRKatDB includes essential gene annotation data, specifically official \textbf{Gene Symbols} and their corresponding \textbf{RefSeq} identifiers.

The current version of miRKatDB encompasses 39,233 unique miRs, 150,349 target genes, and over 6,839,289 miR-target interactions, primarily focused on human data.
Key data fields captured for each interaction within the consolidated miR-target dataset include:

\begin{itemize}
    \item \textbf{miRNA}: The full mature name of the microRNA, as defined by miRBase.
    \item \textbf{mRNA}: The official symbol of the target gene.
    \item \textbf{Score}: A score based on the source database miRDB (ranging from 0 to 1) assigned for each miR-target interaction based on metrics from the source dataset. For experimentally validated interactions (miRTarBase), this score is uniformly set to 1.
    \item \textbf{Source}: The name of the original database from which the interaction data were obtained. There are currently two sources, miRDB and mirTarBase, with 6,831,595 interactions and 7,694, respectively.
\end{itemize}

\subsubsection*{Schema and Architecture}\label{sec:mirkatdb_schema}

miRKatDB was implemented using MySQL version 14.14 as its relational database management system. The foundational architecture was derived from the existing miRBase database structure \cite{Kozomara2019MiRBase:Function, Griffiths-Jones2006MiRBase:Nomenclature}, which was subsequently modified and extended to meet the specific requirements of the miRKat project to integrate a broader range of miR-related data.

During schema design, a primary objective was to maintain the integrity of the original miRBase table structures, thereby ensuring future compatibility and leveraging its established standard for miR information. Consequently, while new tables were introduced and adapted, no existing miRBase tables were fundamentally altered. Key modifications included the addition of new tables to store consolidated and project-specific data. These new tables include, but are not limited to:

\begin{itemize}
    \item \texttt{mirna\_seed}: To store miR seed sequence information, primarily obtained from TargetScan \cite{Agarwal2015PredictingMRNAs}.
    \item \texttt{gene\_mirna}: To house the integrated miR-target interaction data, consolidating entries from various sources like miRDB \cite{Wang2008MiRDB:Interface} and miRTarBase \cite{Huang2022MiRTarBaseInteractions}.

\end{itemize}

\begin{figure}[h!]
    \centering
    \includegraphics[width=.8\linewidth]{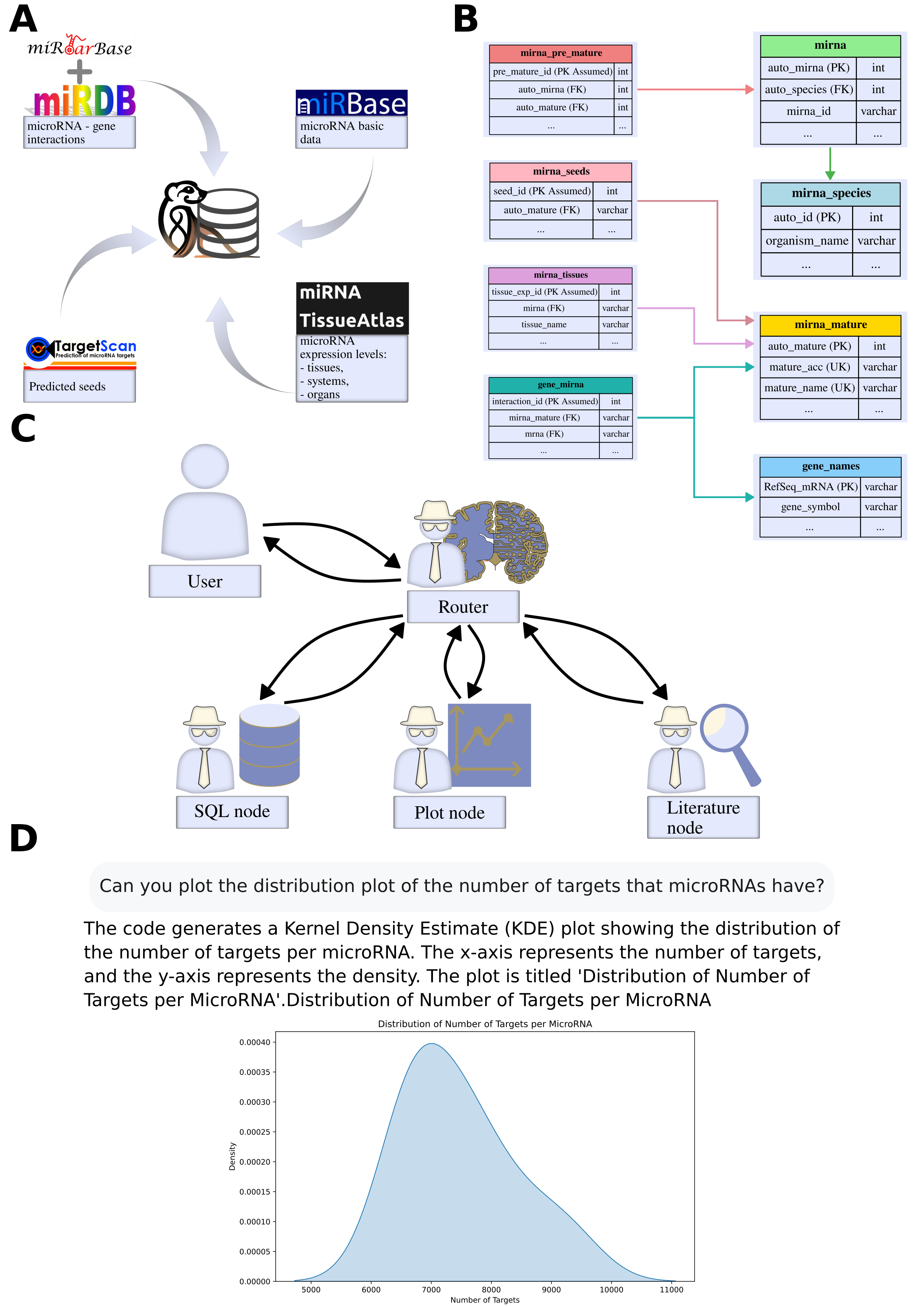}
    \caption{Overview of the miRKat Suite. (A) Data sources of miRKAtDB B) Schematic representation of the miRKatDB, highlighting key tables and their relationships, which integrate data from various public repositories. (C) Diagram illustrating the agent workflow and orchestration within miRKatAI, managed by LangGraph, showing the interaction between the Master Router Agent and specialised agents (SQL, Literature, Plotting). (D) Example of a user interaction with miRKatAI, depicting a natural language query regarding microRNA targets and the corresponding data visualisation answering the question 'can you plot the distribution plot of the number of targets that microRNAs have'.}
    \label{fig:figure}
\end{figure}

The schema was designed to support efficient querying of miR-target relationships, alongside associated gene and miR metadata such as sequences, genomic locations, and miR:target prediction scores, based on miRDB. An overview of the miRKatDB schema, illustrating the key tables and their relationships, is presented in Figure \ref{fig:figure}A,B. Comprehensive documentation for the newly created tables and the original miRBase tables (for which detailed schema information was not readily available) was developed, including table diagrams and field descriptions. This documentation is available in supplementary materials. 
The database server was configured on an Ubuntu Server 20.04 LTS system. Remote access allowed programmatic connections from the miRKatAI agent and other analytical tools. 
Version control for database schemas, configuration scripts, and documentation was maintained using Git, ensuring a reproducible and well-documented development process. The repository is available at \href{https://github.com/GuerreroVazquez/miRKat/tree/dev}{miRKatDB}.

\subsection*{miRKatAI: An AI-Powered Querying and Analysis Agent} \label{sec:mirkatai}
To enhance the accessibility and utility of miRKatDB, we developed miRKatAI. miRKatAI provides a sophisticated natural language interface for miRKatDB and significantly extends its utility by functioning as a dynamic research assistant, adept at integrating information from diverse modalities. It leverages the advanced reasoning of large language models (LLMs) \cite{Team2023Gemini:Models} and the robust orchestration capabilities of the LangGraph framework \cite{Langchain-ai/langgraph:Graphs.}.

miRKatAI allows for seamless translation of natural language questions into precise miRKatDB queries, then augments these findings by performing grounded searches of current scientific literature. Furthermore, miRKatAI can distil complex information into accessible visualisations, enabling users to grasp data trends intuitively. By orchestrating database querying, literature synthesis, and data visualisation within a conversational interface, miRKatAI aims to significantly streamline the research workflow, transforming raw data into actionable insights and accelerating the pace of scientific discovery in the miR field.

\subsubsection*{Technological Framework} \label{sec:mirkatai_framework}

miRKatAI is built using Python, with \textbf{LangGraph} managing the stateful, cyclical workflow between agents.

\subsubsection*{Large Language Models (LLMs)}
A key aspect of miRKatAI is its strategic use of different Google Gemini models, tailored to the specific needs of each agent. This allows for an optimisation of performance, cost, and capability across the system:

\begin{itemize}
    \item \textbf{Master Router Agent}: Employs \texttt{gemini-2.0-flash}, chosen for its speed and efficiency in analysing user requests, understanding intent, and directing the query to the appropriate specialised agent (SQL, Literature, or Plot) 
    \item \textbf{Literature Research Agent}: Also utilises \texttt{gemini-2.0-flash} for processing queries requiring external information and synthesising answers from web search results.
    \item \textbf{SQL Agent}: Leverages \texttt{gemini-2.5-flash-preview}, a more advanced model variant, for translating natural language questions into accurate and efficient SQL queries for miRKatDB.
    \item \textbf{Plotting Agent}: Similarly uses \texttt{gemini-2.5-flash-preview} to generate Python code for data visualisations, requiring precise instruction following and code generation capabilities.
\end{itemize}

\subsubsection*{Agent Workflow and Orchestration}
The \textbf{Master Router Agent} is central to miRKatAI's operation. It first analyses the user's query and the conversation history to determine the next best step. Based on this assessment, it routes the request to one of the specialised agents:

\begin{itemize}
    \item The \textbf{SQL Agent} if structured data from miRKatDB is required.
    \item The \textbf{Literature Research Agent} if the query necessitates broader context, definitions, or recent research findings not present in the database.
    \item The \textbf{Plotting Agent} if the user explicitly requests a data visualisation.
\end{itemize}

When a specialised agent responds, the Master Router evaluates whether the user's original question has received an adequate answer using Gemini \cite{Team2023Gemini:Models}. If it hasn't, the Master Router identifies the additional information or processing needed and may re-engage the same or a different specialised agent. This process continues iteratively until the system formulates a comprehensive answer or reaches a predefined interaction limit. At that point, the Master Router prepares and presents the final answer to the user.
A simplified version of this process is shown in Figure \ref{fig:figure}C.

\paragraph{Specialised Agent Functionalities}
\begin{itemize}
    \item \textbf{SQL Agent}: This agent interacts with the miRKatDB using function calling. It receives detailed metadata about the database schema, including table structures, column descriptions, and relationships. This contextual information enables the LLM to generate more accurate and relevant SQL queries, effectively bridging the gap between natural language and the structured database.
   \item \textbf{Literature Research Agent}: This agent utilises the Google Search API ground web search to find relevant external information. While performing searches, the agent instruction prompts prioritise and synthesise information from sources likely to be peer-reviewed or of high scientific standing, aiming to provide reliable context and current research findings.
    \item \textbf{Plotting Agent}: This agent leverages a code execution tool. When invoked, the LLM generates Python code snippets, primarily using established libraries such as \textbf{Matplotlib} \cite{Hunter2007Matplotlib:Environment} and \textbf{Seaborn} \cite{Waskom2021Seaborn:Visualization}, to create visualisations based on the data provided from previous steps (e.g., SQL query results). The framework is designed to be extensible, allowing for the future addition of more complex plotting capabilities or different visualisation libraries.
\end{itemize}

This technological framework, combining versatile LLMs with a robust agent orchestration system and specialised tools, allows miRKatAI to function as an adaptable and powerful assistant for miR research.

\section*{Results}\label{sec:mirkatai_capabilities}

miRKatAI empowers researchers by offering a suite of intelligent capabilities that streamline and enhance the microRNA research process. Users can leverage miRKatAI to:

\begin{itemize}
    \item \textbf{Perform Complex Database Queries using Natural Language:} miRKatAI translates user questions posed in natural language into precise SQL queries for execution against the miRKatDB. This allows researchers to interrogate the database for specific miR information, such as identifying miRs that target a particular gene or are associated with specific biological pathways, without requiring direct SQL knowledge.

    \item \textbf{Augment Database Findings with Current Literature:} Beyond structured data retrieval, miRKatAI can perform grounded searches of external scientific literature. This capability enables users to obtain up-to-date context, functional explanations, mechanistic insights, or recent research findings related to their queries, effectively bridging the gap between database information and the latest scientific discoveries. The system is guided by prompting to prioritise information from reliable sources.

    \item \textbf{Generate Data Visualisations:} miRKatAI can create basic data visualisations, such as bar charts or scatter plots, based on the information retrieved from miRKatDB or synthesised from literature. This feature allows for a more intuitive understanding of data trends, such as comparing the number of targets for different miRs or visualising expression patterns, directly within the conversational workflow.

    \item \textbf{Engage in Multi-Turn Conversational Interactions:} The system supports iterative dialogue, enabling users to refine their queries, ask follow-up questions, and explore data progressively. This conversational approach allows for a more natural and dynamic interaction, facilitating a deeper exploration of miR data and hypotheses.

    \item \textbf{Simplify Access to Consolidated Information:} By integrating these capabilities, miRKatAI acts as an intelligent interface that simplifies access to consolidated miR information, helping researchers to quickly gather, synthesise, and interpret relevant data, thereby accelerating their research efforts.
    \end{itemize}

These capabilities were validated in a qualitative assessment by two microRNA research experts from the University of Galway. The experts confirmed that the tool's functionalities are relevant to common research scenarios and effectively lower the barrier to complex data exploration and literature synthesis.

\section*{Example Usage}\label{sec:example}

To illustrate the practical application and conversational interface of miRKatAI, we considered a common research scenario in which a user investigates the targets of a specific miRs.
\subsection*{Example 1}
Assuming the user is curious about the number of targets of a selected miR (s), miRKatAI will first establish how many targets each miR has. For example, a query can be in the form: `Can you plot the distribution of the number of targets that microRNAs have?’. miRKatAI will process this request by first employing its natural language understanding capabilities. The Master Router Agent will identify the need for database querying to fetch the target gene information and the subsequent requirement for data visualisation. It will then direct the SQL Agent to construct and execute a precise query against miRKatDB, retrieving the number of targets per miR. From these, follow-up information on which miR has the most targets will be generated through searching the literature  (through Ground search; Figure 1D).

miRKatAI displays a user’s natural language query entered into the miRKatAI interface, followed by the system’s response. This response will typically consist of a textual summary and a graphical representation. In this example, the query about the distribution of the number of targets for microRNAs will initiate the system analysis of the miRKatDB to determine the number of targets each microRNA has, and generates a plot of this distribution. Additionally, a follow-up question to identify which microRNA has the most targets will also be supported by a literature search.

Figure \ref{fig:figure}D illustrates this interaction. It displays a user's natural language query entered into the miRKatAI interface, followed by the system’s response. This response typically consists of a textual summary and a graphical representation. In this example, the user inquires about the distribution of the number of targets for microRNAs. The system analyses the database to determine the number of targets each microRNA has and then generates a plot of this distribution. 

\subsection*{Example 2}

Another example use case is of the user asking: `What is the role of miR-143 in ageing?', for which miRKatAI response is shown below:

\begin{shadedquotation}
\begin{quote}
\small
Based on recent peer-reviewed literature, here's a summary of the role of miR-143 in aging:
\textbf{General Role and Expression:}
\begin{itemize}
    \item miR-143-3p is highly expressed in skeletal muscle tissue.[\href{https://vertexaisearch.cloud.google.com/grounding-api-redirect/AUZIYQGiY5Ux3ywN_F7HqJcQdEqJfQSW4Pw_BEbcLy82GJ3mDA_TFMWkvW5ouBgd9fhMQWlC78IpXF9BZuqJ7YZx8Y35w9P601cyVlJ-VHywauw5K1ntKFYeaMajrrmNt21sw70LZsXluR04FUAHnrY1a1Nvz1TK3RFUpd_fI5jOxFuW4_s_pPgtNGqAxeOKSFmKXtK0y4pX8V0Sda4LT1mzxEC-DpBNxq0guPGs9kaS7dEPXmBZmufmqaOfuZeAhQ2ceKoYCyA=}{1}] Its concentration decreases during aging and is involved in insulin signaling.[\href{https://vertexaisearch.cloud.google.com/grounding-api-redirect/AUZIYQGiY5Ux3ywN_F7HqJcQdEqJfQSW4Pw_BEbcLy82GJ3mDA_TFMWkvW5ouBgd9fhMQWlC78IpXF9BZuqJ7YZx8Y35w9P601cyVlJ-VHywauw5K1ntKFYeaMajrrmNt21sw70LZsXluR04FUAHnrY1a1Nvz1TK3RFUpd_fI5jOxFuW4_s_pPgtNGqAxeOKSFmKXtK0y4pX8V0Sda4LT1mzxEC-DpBNxq0guPGs9kaS7dEPXmBZmufmqaOfuZeAhQ2ceKoYCyA=}{1}]
    \item Dysregulation of miR-143 expression is relevant in regenerating muscle of older individuals.[\href{https://vertexaisearch.cloud.google.com/grounding-api-redirect/AUZIYQGiY5Ux3ywN_F7HqJcQdEqJfQSW4Pw_BEbcLy82GJ3mDA_TFMWkvW5ouBgd9fhMQWlC78IpXF9BZuqJ7YZx8Y35w9P601cyVlJ-VHywauw5K1ntKFYeaMajrrmNt21sw70LZsXluR04FUAHnrY1a1Nvz1TK3RFUpd_fI5jOxFuW4_s_pPgtNGqAxeOKSFmKXtK0y4pX8V0Sda4LT1mzxEC-DpBNxq0guPGs9kaS7dEPXmBZmufmqaOfuZeAhQ2ceKoYCyA=}{1},\href{https://vertexaisearch.cloud.google.com/grounding-api-redirect/AUZIYQHshNCKEwd9uU2jT8J_PfCczk7cmaxIqa-o__MbqIP_tGogxrYtVOUc99E_7zyjMBKjp7aWLuYXBu6y_uifiYnTCVjlUtx6v9ncgklOq2kakkKzinGERC8JDsY3i3PjbZvM5TlX}{2}]
    \end{itemize}
\textbf{Muscle Aging and Regeneration:}
\begin{itemize}
    \item Downregulation of miR-143 in myogenic cells from old mice is associated with increased cell viability.[\href{https://vertexaisearch.cloud.google.com/grounding-api-redirect/AUZIYQGiY5Ux3ywN_F7HqJcQdEqJfQSW4Pw_BEbcLy82GJ3mDA_TFMWkvW5ouBgd9fhMQWlC78IpXF9BZuqJ7YZx8Y35w9P601cyVlJ-VHywauw5K1ntKFYeaMajrrmNt21sw70LZsXluR04FUAHnrY1a1Nvz1TK3RFUpd_fI5jOxFuW4_s_pPgtNGqAxeOKSFmKXtK0y4pX8V0Sda4LT1mzxEC-DpBNxq0guPGs9kaS7dEPXmBZmufmqaOfuZeAhQ2ceKoYCyA=}{1}]
    \item miR-143-3p regulates myoblast senescence, myogenesis in vitro, and muscle regeneration.[\href{https://vertexaisearch.cloud.google.com/grounding-api-redirect/AUZIYQHshNCKEwd9uU2jT8J_PfCczk7cmaxIqa-o__MbqIP_tGogxrYtVOUc99E_7zyjMBKjp7aWLuYXBu6y_uifiYnTCVjlUtx6v9ncgklOq2kakkKzinGERC8JDsY3i3PjbZvM5TlX}{2},\href{https://vertexaisearch.cloud.google.com/grounding-api-redirect/AUZIYQGiY5Ux3ywN_F7HqJcQdEqJfQSW4Pw_BEbcLy82GJ3mDA_TFMWkvW5ouBgd9fhMQWlC78IpXF9BZuqJ7YZx8Y35w9P601cyVlJ-VHywauw5K1ntKFYeaMajrrmNt21sw70LZsXluR04FUAHnrY1a1Nvz1TK3RFUpd_fI5jOxFuW4_s_pPgtNGqAxeOKSFmKXtK0y4pX8V0Sda4LT1mzxEC-DpBNxq0guPGs9kaS7dEPXmBZmufmqaOfuZeAhQ2ceKoYCyA=}{1}]
    \item Disrupted expression of miR-143-3p and its target gene, \textit{Igfbp5}, plays a role in muscle regeneration.[\href{https://vertexaisearch.cloud.google.com/grounding-api-redirect/AUZIYQHshNCKEwd9uU2jT8J_PfCczk7cmaxIqa-o__MbqIP_tGogxrYtVOUc99E_7zyjMBKjp7aWLuYXBu6y_uifiYnTCVjlUtx6v9ncgklOq2kakkKzinGERC8JDsY3i3PjbZvM5TlX}{2}]
    \item Downregulation of miR-143 during aging might be a compensatory mechanism to improve myogenesis; however, increased levels of \textit{Igfbp5} (a miR-143 target gene) can increase cell senescence, which affects myogenesis.[\href{https://vertexaisearch.cloud.google.com/grounding-api-redirect/AUZIYQHshNCKEwd9uU2jT8J_PfCczk7cmaxIqa-o__MbqIP_tGogxrYtVOUc99E_7zyjMBKjp7aWLuYXBu6y_uifiYnTCVjlUtx6v9ncgklOq2kakkKzinGERC8JDsY3i3PjbZvM5TlX}{2}]
    \end{itemize}
\textbf{Age-Related Diseases:}
\begin{itemize}
    \item In cartilage aging and osteoarthritis, the expression of miR-143-3p decreases after IL-1 $\beta$ treatment in human primary chondrocytes.[\href{https://vertexaisearch.cloud.google.com/grounding-api-redirect/AUZIYQF8ilCDOJWZiymj5a7M5MSRr6C197e7_riLSu9jARstlh1xyIqrYnvBZqEkcofifiblbUAIaicldRH8Fw6gKBwyX3c8vWzXzV6s2fV4TLvlGtA5jofEV4yjLS0fqQ==}{3}]
    \item MiR-143/145 levels rise as estrogen levels drop in postmenopausal women.[\href{https://vertexaisearch.cloud.google.com/grounding-api-redirect/AUZIYQFkzpI0n3bILlq_N2Nf83udhPGQjcg7RuDPp8pfLbX0NSP8BUd9TsvjrNKrBr4GvY-wxMrjhuO7qiBUnQS9dIJO9j61YxrsEpkWHizLlqLW_mgg1UXM38Wb6jhp-WgvitJa5HK6CmfzhotAeWYhx0yyiyyxxV03aKs8J6xUMS_lFw==}{4}]
    \item Dysregulation of circulating microRNA (c-miRNA), including miR-143-3p, with aging may serve as a predictive marker and reflect molecular mechanisms causing declines in muscle mass, increased fat mass, and anabolic resistance.[\href{https://vertexaisearch.cloud.google.com/grounding-api-redirect/AUZIYQFkzpI0n3bILlq_N2Nf83udhPGQjcg7RuDPp8pfLbX0NSP8BUd9TsvjrNKrBr4GvY-wxMrjhuO7qiBUnQS9dIJO9j61YxrsEpkWHizLlqLW_mgg1UXM38Wb6jhp-WgvitJa5HK6CmfzhotAeWYhx0yyiyyxxV03aKs8J6xUMS_lFw==}{5}]
    \end{itemize}
\textbf{Other Age-Related Effects:}
\begin{itemize}
    \item Circulating miR-143-3p levels are predictive of aging.[\href{https://vertexaisearch.cloud.google.com/grounding-api-redirect/AUZIYQFkzpI0n3bILlq_N2Nf83udhPGQjcg7RuDPp8pfLbX0NSP8BUd9TsvjrNKrBr4GvY-wxMrjhuO7qiBUnQS9dIJO9j61YxrsEpkWHizLlqLW_mgg1UXM38Wb6jhp-WgvitJa5HK6CmfzhotAeWYhx0yyiyyxxV03aKs8J6xUMS_lFw==}{5}]
    \item miR-143 is involved in the development of cisplatin resistance via \textit{IGF1R} and \textit{BCL2}.[\href{https://vertexaisearch.cloud.google.com/grounding-api-redirect/AUZIYQEukauZodKv39HMdWR4hwflMx5reoE3Z1lwbpl9vjfi5FBd-PQLKeNXrI90CvmLQuSapHGfzEo-Reb4iqKfI31_I-DFcPStYalneEfLyezKIchfUyexF2-93n_jmE480vXxzf5K}{6}]
    \item In an Alzheimer's disease cell model, miR-143-3p inhibition promotes neuronal survival by targeting \textit{neuregulin-1} (\textit{NRG1}).[\href{https://vertexaisearch.cloud.google.com/grounding-api-redirect/AUZIYQHlZvAsdChmkpuA5Z0EqFlml6fSSiMyibHGk8ocIBx1FCoJV7YFMA7DmSIo6E0AhV6hEqKOYTFje4CgtTnA7KdRae-TYqnKjDgyT5UKUfw-x_WHTY2thE7RIMZ5pRd3Fy0stgALfM_TyR7EN-bqWjGhdPuJvFXcZQpOM3zbl-cwg-eRC1FBGJMoGcc58GzBfDBArqb_yr19ylCXGMBcoL8Z7PSQIkDmt_6veMKrkG-VUEkFBigsgkPnXWrOslbm30YrX30K7OM79ZGTL1mtQapCYG8fbaM=}{7}]
    \end{itemize}
\textbf{References:}

\begin{enumerate}
    \item \href{https://vertexaisearch.cloud.google.com/grounding-api-redirect/AUZIYQGiY5Ux3ywN_F7HqJcQdEqJfQSW4Pw_BEbcLy82GJ3mDA_TFMWkvW5ouBgd9fhMQWlC78IpXF9BZuqJ7YZx8Y35w9P601cyVlJ-VHywauw5K1ntKFYeaMajrrmNt21sw70LZsXluR04FUAHnrY1a1Nvz1TK3RFUpd_fI5jOxFuW4_s_pPgtNGqAxeOKSFmKXtK0y4pX8V0Sda4LT1mzxEC-DpBNxq0guPGs9kaS7dEPXmBZmufmqaOfuZeAhQ2ceKoYCyA=}{researchgate.net}
    \item \href{https://vertexaisearch.cloud.google.com/grounding-api-redirect/AUZIYQHshNCKEwd9uU2jT8J_PfCczk7cmaxIqa-o__MbqIP_tGogxrYtVOUc99E_7zyjMBKjp7aWLuYXBu6y_uifiYnTCVjlUtx6v9ncgklOq2kakkKzinGERC8JDsY3i3PjbZvM5TlX}{nih.gov}
    \item \href{https://vertexaisearch.cloud.google.com/grounding-api-redirect/AUZIYQEBKjVXsFLOy2w_tyW_5p1rlvmhAHE6IA00MZoeo_kMYFC88qRJP8FytnHDU6a3nwkGjKHcrIEbWtZaVOWjODUmic-039wfZOufW1X3dSJ1ZVxKQJzcmwZnhotcF8D-3s9KxNIbCGpqQubsO7Zz23RDcqbtfVtuCEsLG9JH-oT8zVECkYxy0eWURSZobr7o}{frontiersin.org}
    \item \href{https://vertexaisearch.cloud.google.com/grounding-api-redirect/AUZIYQF8ilCDOJWZiymj5a7M5MSRr6C197e7_riLSu9jARstlh1xyIqrYnvBZqEkcofifiblbUAIaicldRH8Fw6gKBwyX3c8vWzXzV6s2fV4TLvlGtA5jofEV4yjLS0fqQ==}{thno.org}
    \item \href{https://vertexaisearch.cloud.google.com/grounding-api-redirect/AUZIYQFkzpI0n3bILlq_N2Nf83udhPGQjcg7RuDPp8pfLbX0NSP8BUd9TsvjrNKrBr4GvY-wxMrjhuO7qiBUnQS9dIJO9j61YxrsEpkWHizLlqLW_mgg1UXM38Wb6jhp-WgvitJa5HK6CmfzhotAeWYhx0yyiyyxxV03aKs8J6xUMS_lFw==}{oup.com}
    \item \href{https://vertexaisearch.cloud.google.com/grounding-api-redirect/AUZIYQEukauZodKv39HMdWR4hwflMx5reoE3Z1lwbpl9vjfi5FBd-PQLKeNXrI90CvmLQuSapHGfzEo-Reb4iqKfI31_I-DFcPStYalneEfLyezKIchfUyexF2-93n_jmE480vXxzf5K}{mdpi.com}
    \item \href{https://vertexaisearch.cloud.google.com/grounding-api-redirect/AUZIYQHlZvAsdChmkpuA5Z0EqFlml6fSSiMyibHGk8ocIBx1FCoJV7YFMA7DmSIo6E0AhV6hEqKOYTFje4CgtTnA7KdRae-TYqnKjDgyT5UKUfw-x_WHTY2thE7RIMZ5pRd3Fy0stgALfM_TyR7EN-bqWjGhdPuJvFXcZQpOM3zbl-cwg-eRC1FBGJMoGcc58GzBfDBArqb_yr19ylCXGMBcoL8Z7PSQIkDmt_6veMKrkG-VUEkFBigsgkPnXWrOslbm30YrX30K7OM79ZGTL1mtQapCYG8fbaM=}{termedia.pl}
    \end{enumerate}
\end{quote}
\end{shadedquotation}

The links redirect to the journal where the information was obtained. 

These examples demonstrate how miRKatAI effectively combines its capabilities to translate a user's scientific question into actionable insights. It simplifies data exploration and supports hypothesis generation directly within an intuitive conversational workflow, integrating natural language querying, direct database access, and on-demand visualisation.

\section*{Availability} \label{sec:availability}

The miRKat Suite, comprising miRKatDB and miRKatAI (currently version 1.0), is freely available to the research community to facilitate microRNA studies.
\textbf{miRKatAI} can be accessed via its web interface at \href{https://mirkat.ie/}{miRKat.ie}.
The source code, database schema details, and additional documentation for the miRKat Suite are intended to be made available through a public Git repository \href{https://github.com/PasTaco/miRKatAI} to facilitate transparency, community contributions, and local deployment.

We encourage researchers to utilise these resources and welcome feedback for future development and improvements.

\section*{Conclusions} \label{sec:conclusion}

This study introduces miRKatAI, a multi-agent that integrates the comprehensive miRKatDB, representing a step towards simplifying and accelerating microRNA research. It provides a unified platform for natural language database querying, grounded literature synthesis, and on-demand data visualisation. miRKatAI lowers the technical barriers to complex data exploration, empowering researchers to focus on scientific discovery


The challenges of data integration, analysis, and staying current with literature are not unique to microRNA research. In response, a new paradigm is emerging across scientific domains: the use of AI-powered agents to assist researchers. In the biological sciences, these systems are being developed to automate complex workflows and democratise access to sophisticated analytical tools. For instance, systems like BioAgents \cite{Mehandru2025BioAgents:Systems} and the recently proposed BioMaster \cite{Su2025BioMaster:Workflow} employ multi-agent architectures to assist researchers in executing multi-step bioinformatics pipelines, ranging from RNA-seq analysis to genome assembly, often through natural language commands. In the related field of chemistry, the ChemCrow agent has demonstrated the ability to orchestrate a diverse set of computational tools to assist in tasks ranging from literature synthesis to planning chemical reactions \cite{Bran2023ChemCrow:Tools}. These advancements underscore the potential for intelligent agents to serve as powerful research assistants, capable of bridging the gap between complex research questions and the computational steps necessary to address them. It is within this innovative context that we have developed the miRKat Suite.

miRKatAI is presented as a dynamic tool with considerable potential for continuous improvement and expansion. 
miRKatAI offers a valuable assistant to researchers navigating the intricate landscape of microRNA biology, facilitating the generation of new hypotheses and uncovering novel regulatory mechanisms.

\section*{Limitations and Future Directions}
While miRKatAI represents a significant step towards simplifying miR research, it is important to acknowledge its current limitations. Presently, the system's capabilities are focused on data retrieval, synthesis, and basic visualisation; it does not perform novel statistical analyses or complex bioinformatics computations. Furthermore, miRKatAI is designed to query its integrated database and cannot currently accept or analyse user-uploaded datasets.

As with any system built upon large language models, its responses are not entirely deterministic and can vary between identical queries. The performance and capabilities of miRKatAI are also intrinsically linked to the underlying Google Gemini models, making it dependent on their updates and availability. Although we employ grounding techniques to minimise inaccuracies, the potential for factual errors or `hallucinations' remains, necessitating critical evaluation of all outputs by the user. For optimal results, users must still formulate their queries with care, and response times can vary depending on query complexity and system load.

We present miRKatAI's first generation iteration. Future work will focus on addressing these limitations by incorporating more advanced analytical tools, exploring secure pathways for user data integration, and enhancing the determinism and speed of the agentic responses. We believe that community feedback will be invaluable in guiding its evolution into an even more powerful and reliable research assistant.


\section*{List of Abbreviations}
\begin{itemize}
    \item[\textbf{AI}] Artificial Intelligence 
    \item[\textbf{LLM}] Large Language Model
    \item[\textbf{miR}] microRNA 
\end{itemize}

\section*{Declarations}

\subsubsection*{Ethics approval and consent to participate}
Not applicable.

\subsubsection*{Consent for publication}
Not applicable.

\subsubsection*{Availability of data and materials}
Code for miRKatDB is published on GitHub in the repository  \href{https://github.com/GuerreroVazquez/mirKat}{GuerreroVazquez/mirKat}.
Code for miRKatAI is published on GitHub in the repository
\href{https://github.com/PasTaco/MirKatAI}{PasTaco/MirKatAI}.

\subsubsection*{Competing interests}
The authors have no competing interests to declare.

\subsubsection*{Funding}
KW, KGV, JUV, ES and POB are funded by Research Ireland, 18/CRT/6214; KW is funded by Research Ireland SFI19/FFP/6709.

\subsubsection*{Authors' contributions}
\begin{itemize}
    \item KGV and JUV contributed to the concept, miRKatAI tool generation, and manuscript preparation.  
\item KGV and KW created the miRKatDB database.  
\item KW contributed funding, concept development, manuscript writing, and miRKatAI assessment.  
\item POB contributed funding and manuscript writing.  
\item ES contributed funding and manuscript writing. 
\end{itemize}

All authors revised the manuscript.

\subsubsection*{Acknowledgements}
We want to thank Dr. Brian McDonagh, University of Galway, for help with the qualitative assessment of the tool.

\section*{Availability and Requirements}

\begin{description}
    \item[Project name:] miRKatAI
    \item[Project home page:] \url{https://github.com/PasTaco/MirKatAI}, \url{https://mirkat.ie}
    \item[Operating system(s):] Platform independent
    \item[Programming language:] Python
    \item[Other requirements:] Accessing the hosted web service at \url{http://mirkat.ie/} requires only a modern web browser. For local installation and deployment from the source code, the requirements include: Python 3.12 or higher, uv, Google Cloud SDK, Terraform, and Make. All other Python package dependencies are detailed in the project repository.
    \item[License:] MIT License
    \item[Any restrictions to use by non-academics:] None
\end{description}

\bibliographystyle{ieeetr}
\bibliography{references}

\end{document}